\documentclass[graybox]{svmult}

\usepackage[utf8]{inputenc}
\usepackage[T1]{fontenc}
\usepackage{xcolor}
\definecolor{blue}{RGB}{0,0,153}
\usepackage[%
  colorlinks=true,
  allcolors=blue
]{hyperref}
\usepackage{url}
\usepackage{enumitem}
\usepackage{amsfonts}
\usepackage{amssymb}
\usepackage{amsmath}
\usepackage{array}
\usepackage{adjustbox}
\usepackage{mathtools}
\usepackage[ruled,vlined]{algorithm2e}
\usepackage{lmodern}
\usepackage[normalem]{ulem}
\usepackage{makecell}

\usepackage[style=numeric-comp,sorting=none,doi=false,url=false,giveninits,backend=bibtex]{biblatex}
\begin{document}

\title*{
    Illuminating Protein Dynamics: A Review of Computational Methods for Studying Photoactive Proteins
}
\titlerunning{Illuminating Protein Dynamics}

\author{
    Sylwia Czach,\orcidID{0000-0002-1546-1075} \\
    Jakub Rydzewski,\orcidID{0000-0003-4325-4177} and\\
    Wiesław Nowak\orcidID{0000-0003-2584-1327}
}

\institute{
    Sylwia Czach, Jakub Rydzewski, and Wiesław Nowak \at
    Institute of Physics, 
    Faculty of Physics, Astronomy and Informatics\\
    Nicolaus Copernicus University\\
    Grudziadzka 5, 87-100 Toru\'n, Poland\\
    \email{wiesiek@fizyka.umk.pl}
}

\maketitle

\abstract{
Photoactive proteins absorb light and undergo structural changes that enable them to perform essential biological functions. These proteins are critical for understanding light-induced biological processes, making them important in biophysics, biotechnology, and medicine. One effective approach to uncovering photoactive processes is through computational methods. These techniques provide atomic-level insights into the structural, electronic, and dynamic changes that occur upon light absorption. By employing these methods, we can gain a better understanding of processes that are challenging to capture experimentally, such as chromophore isomerization and protein conformational changes. Here, we provide a brief overview of the different families of photoactive proteins and the computational methods used to study them, including bioinformatics, molecular dynamics, and enhanced sampling. Our review can serve as an introduction to computational methods for studying light-activated molecular processes, specifically targeting researchers beginning their journey in this field.
}

\section{Introduction}
Photoactive proteins are specialized to detect and respond to light because they contain a light-sensitive chromophore. By absorbing light at various wavelengths, photoactive proteins enable precise biological activation, enabling them to overcome high energy barriers, leading to conformational changes that rearrange their structure. Many photoactive proteins also demonstrate reversible activation, meaning they can return to their original state after performing their function, allowing them to respond to light repeatedly. These unique characteristics highlight the significance of photoactive proteins in light-dependent biological processes and their potential applications in biotechnology, medicine, and optogenetics~\cite{hughes1997prokaryotic,jiang1999bacterial,hellingwerf2003photoactive,van2004photoreceptor,mroginski2021frontiers}. Many metabolic processes that are critical for the proper functioning of all living organisms proceed via pathways that depend on the interaction of biological macromolecules with light. Examples of such processes are photosynthesis, pigmentation and phototaxis in prokaryotes and plants, vision and circadian regulation, light-dependent gene expression, and DNA repair. Light-regulated proteins are also responsible for the aggregation of cells and the formation of colonies in bacteria~\cite{capece2013small,gozem2017theory,purcell2007photosensory}.

\begin{figure}[p]
  \centering
  \includegraphics[width=1.0\textwidth]{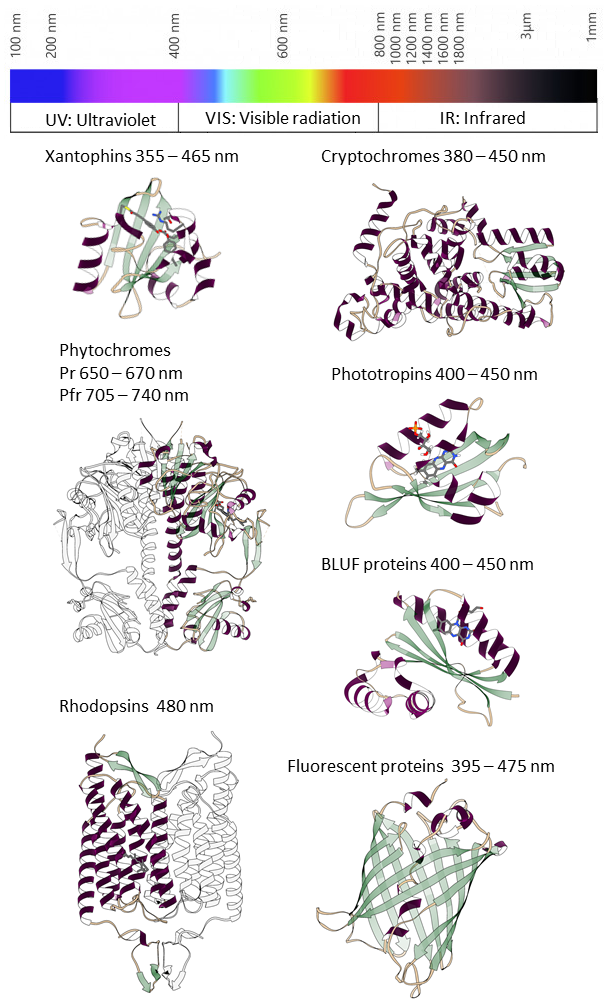}
  \caption{Visible light spectrum and visualization of the structures of example proteins from each family with information on the wavelength to which the chromophore reacts (xantophins PDB:1TS6~\cite{ihee2005visualizing}, cryptochromes PDB:5T5X~\cite{michael2017formation}, phytochromes PDB:4O0P~\cite{takala2014signal}, phototropins PDB:8QI8~\cite{gotthard2024capturing}, BLUF proteins PDB:1X0P~\cite{kita2005structure}, rhodopsins PDB:3UG9~\cite{kato2012crystal}, fluorescent proteins PDB:1GFL~\cite{yang1996molecular}).}
  \label{fig:spectrum}
\end{figure}

\begin{table}
\centering
\caption{Comprehensive classification and overview of photoactive protein families: presentation of protein families, chromophore types, representative examples of chromophores, and photochemical mechanisms underlying light-driven biological processes.}
\label{table:prot}
\begin{tabular}{c|c|c|c}
Protein family & Chromophore & Chromophore example & Mechanism\\ 
\hline
Xantophins & Polyenes 
    & 
    \begin{tabular}{c}
    $p$-Coumaric acid\\[3pt]
    \adjustbox{valign=c}{\includegraphics[width=2.5cm]{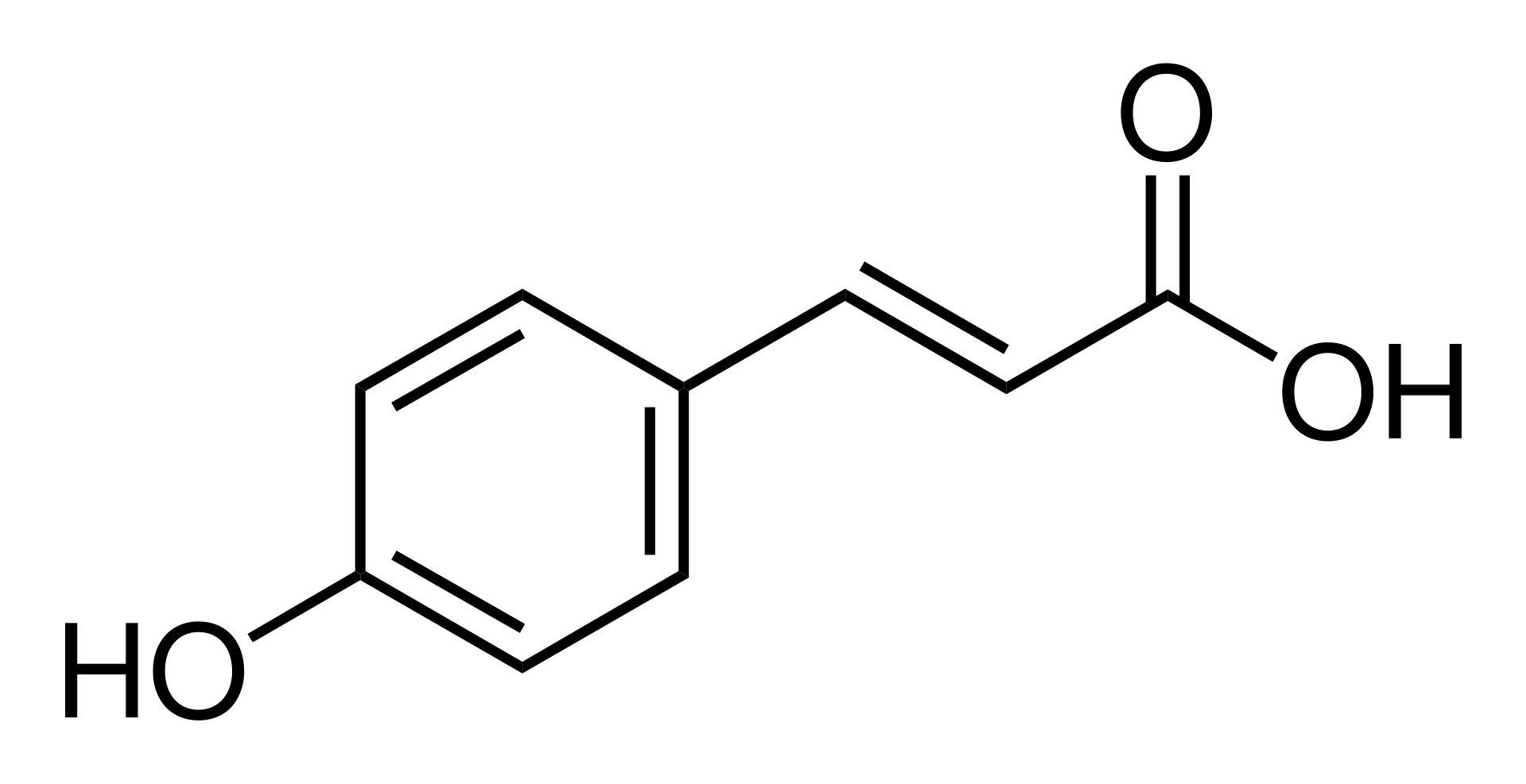}}
    \end{tabular}               
    & 
    \makecell{Trans--cis \\photoisomerization} \\ 
\hline
Phytochromes & Tetrapyroles 
    &
    \begin{tabular}{cc} 
    Biliverdin \\[3pt]
    \adjustbox{valign=c}{\includegraphics[width=3.4cm]{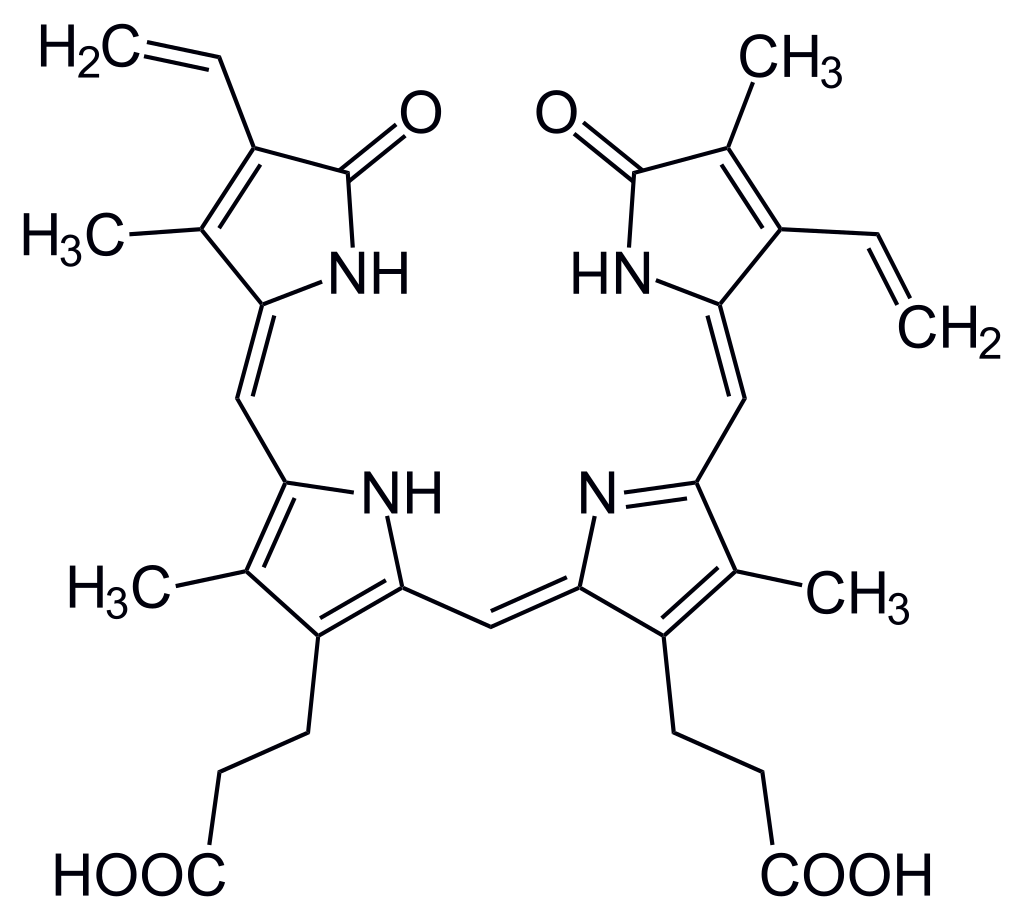}}
    \end{tabular}
    &
    \makecell{Trans--cis \\photoisomerization} \\ 
\hline
Rhodopsins & Polyenes
    & 
    \begin{tabular}{cc}
    Retinal \\[3pt]
    \adjustbox{valign=c}{\includegraphics[width=3cm]{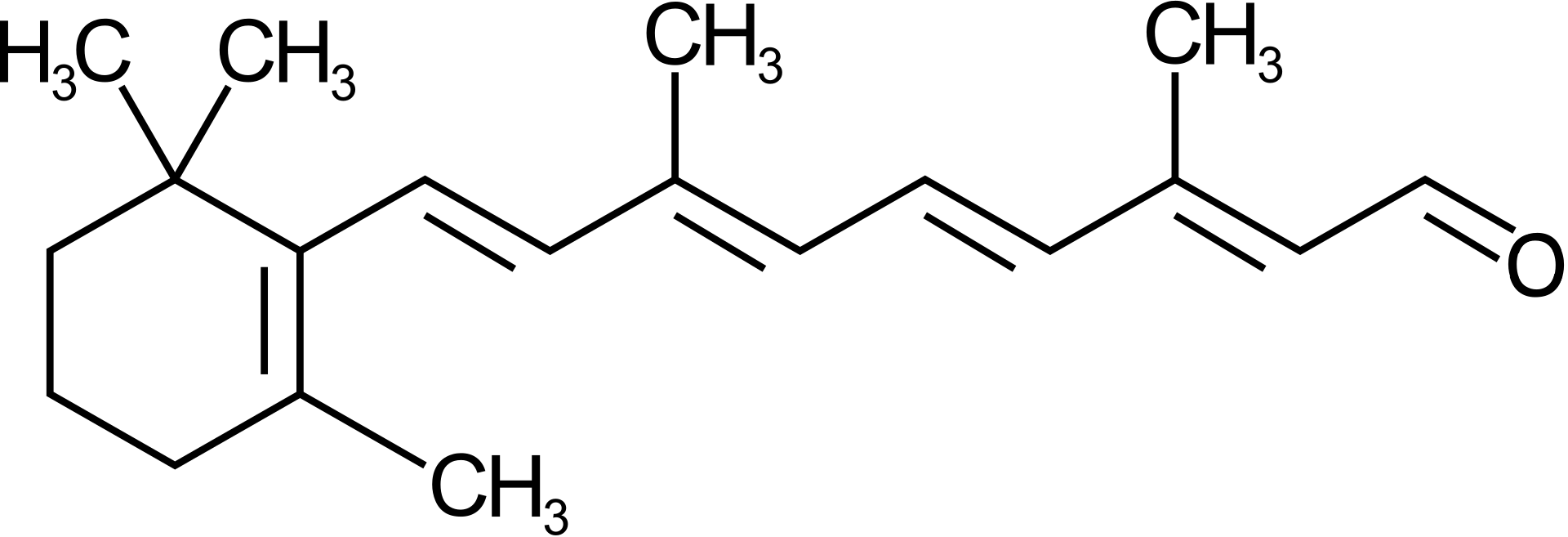}}
    \end{tabular}                         
    & 
    \makecell{Trans--cis \\photoisomerization} \\ 
\hline
Cryptochromes & Aromatic                                                   
    & 
    \begin{tabular}{cc}
    Flavin adenine dinucleotide \\[3pt] 
    \adjustbox{valign=c}{\includegraphics[width=2.5cm,angle=90]{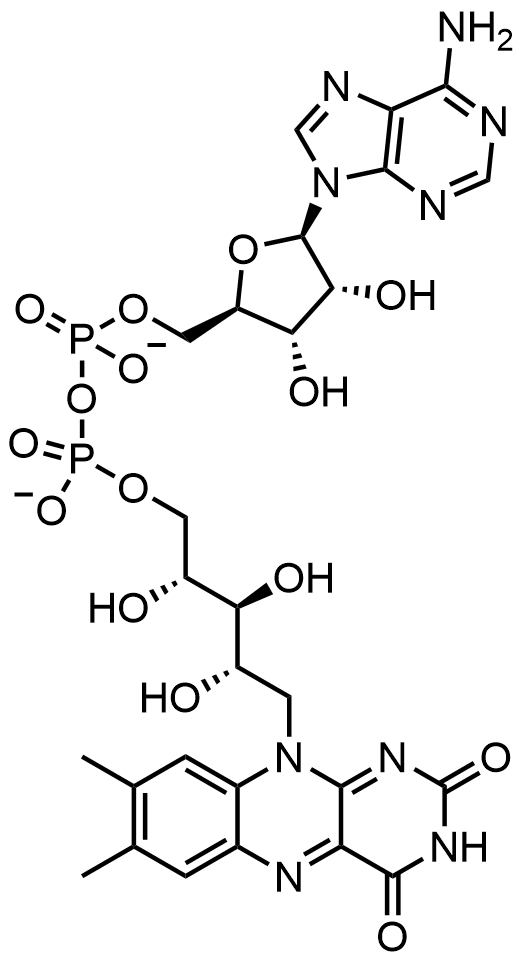}}
    \end{tabular}   
    & 
    Electron transfer \\ 
\hline
Phototropins & Aromatic                                                   
    & 
    \begin{tabular}{cc}
    Flavin adenine mononucleotide \\[3pt] 
    \adjustbox{valign=c}{\includegraphics[width=2.6cm]{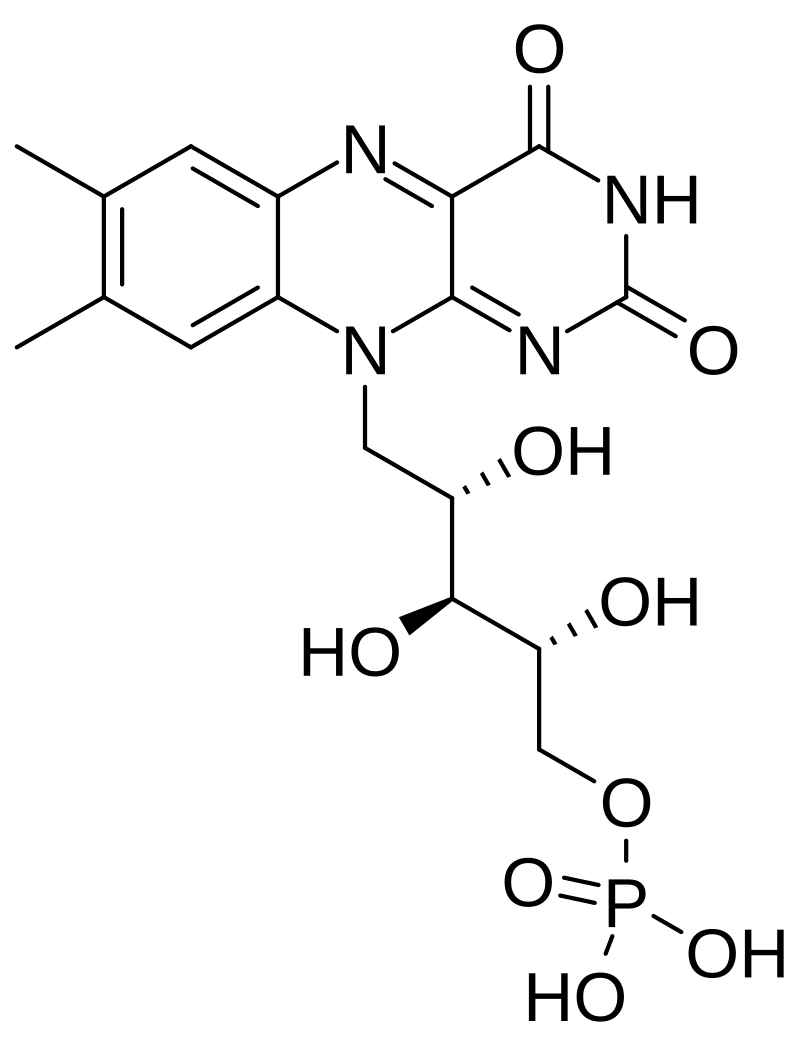}}
    \end{tabular} 
    & 
    \makecell{Cysteinyl adduct \\formation} \\ 
\hline
BLUF proteins & Aromatic 
    & 
    \begin{tabular}{cc}
    Flavin adenine dinucleotide \\[3pt] 
    \adjustbox{valign=c}{\includegraphics[width=2.5cm,angle=90]{fig/fad.png}}
    \end{tabular}   
    & 
    Proton transfer \\ 
\hline
Fluorescent proteins & Amino acids 
    & 
    \begin{tabular}[c]{cc}
    Serine, tyrosine, glycine \\[3pt] 
    \adjustbox{valign=c}{\includegraphics[width=2.9cm,angle=90]{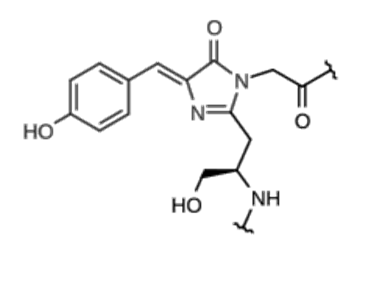}}
    \end{tabular}     
    & 
    \makecell{Fluorophore \\formation} \\
\end{tabular}
\end{table}

Our understanding of photoreceptors has largely been based on structural and spectroscopic data~\cite{moglich2009structure,kennis2013molecular,salvadori2024deciphering}. Because of high-resolution crystallography and cryo-electron microscopy, we have access to detailed structures of many photoreceptors. Additionally, time-resolved spectroscopy enables us to monitor the dynamics of processes triggered by light. Nevertheless, a complete microscopic characterization of the photoactivation process of many photoreceptors remains incomplete. This stems mainly from the difficulties in experimentally reconstructing photoactivation across various temporal scales, ranging from ultrafast femtosecond and picosecond scales of the photochemical event to the slower millisecond scales of protein conformational transitions. Capturing all ultrafast intermediates and understanding interactions between chromophore dynamics and protein conformational changes is far from trivial. Deciphering light-induced reactions is challenging mainly because they occur at the excited states~\cite{groenhof2020observe}.

Addressing these shortcomings, which hinder new technological breakthroughs, can be achieved by studying these photoactive systems through atomistic computational methods~\cite{mroginski2021frontiers}. Methods that have been developed to explore organic photoactive systems include molecular dynamics (MD)~\cite{gozem2017theory}, hybrid quantum-classical (QM/MM) MD~\cite{brunk2015mixed,andruniow2021qm}, enhanced sampling~\cite{bernardi2015enhanced,valsson2016enhancing,bussi2020using,henin2022enhanced}. Currently, there is a need to incorporate machine learning (ML) methods into such techniques~\cite{wang2020machine,rydzewski2023manifold,mehdi2024enhanced,gokdemir2025machine}. The limitations of current computational techniques in predicting dynamics across multiple timescales, combined with the need for a deeper understanding of photoactive proteins in their excited states, are driving the rapid development of new computational methods.

This review aims to introduce the reader to photoactive proteins, their mechanism of action, and their dynamics in response to light. It also provides an overview of computational methods used to study the behavior of photoactive proteins. The review is divided into two main parts. In the first part, six families of photoactive proteins are introduced~\cite{van2004photoreceptor}: xanthopsins~\cite{kort1996xanthopsins,hoff1999global}, phytochromes~\cite{quail1998phytochrome}, rhodopsins~\cite{hoff1997molecular,spudich2000retinylidene}, cryptochromes~\cite{ahmad1993hy4}, phototropins~\cite{huala1997arabidopsis}, and BLUF (blue light using flavins) proteins~\cite{gomelsky2002bluf}. We put special emphasis on the mechanism of their photoactivation, their biological role, and their application in biotechnology and medicine. In addition, the most important structural and functional features relevant to computational studies are presented. The next part of the article consists of a review of computational methods used to study photoactive systems. A brief description of each method is presented, as well as an example of how the method is used to study the proteins presented in the first part. Finally, we discuss the limitations of the described methods, the need to integrate experimental data with computational results, and future directions for computational methods. 

\begin{figure}[b]
  \centering
  \includegraphics[width=1\textwidth]{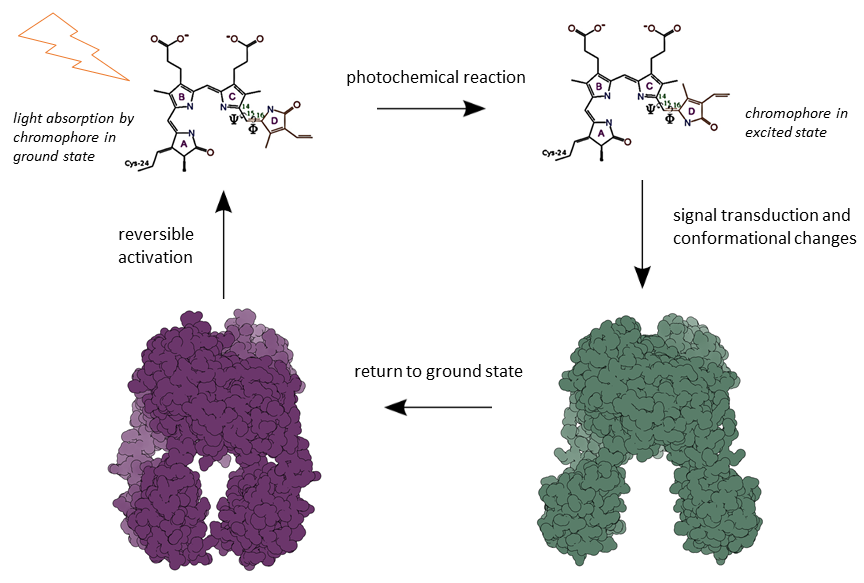}
  \caption{Schematic representation of the action of photoactive proteins.}
  \label{fig:photoproteins}
\end{figure}

\section{Overview of Photoactive Proteins}
While proteins are often sensitive and can be damaged by electromagnetic radiation, photoactive proteins show adaptation to light environmental conditions. Photoactive proteins are specialized biological molecules that act as molecular switches, responding to light by undergoing conformational or chemical changes. Contain light-absorbing chromophores that are critical for organisms to adapt to their light environment, inducing structural changes that drive downstream biochemical signaling pathways~\cite{hughes1997prokaryotic,jiang1999bacterial,hellingwerf2003photoactive,van2004photoreceptor,mroginski2021frontiers}. 
  
A small number of groups can be identified among various photoresponsive proteins. Table~\ref{table:prot} shows the emerging classification of the proteins reported in the literature into six families: xanthopsins~\cite{kort1996xanthopsins,hoff1999global}, phytochromes~\cite{quail1998phytochrome}, rhodopsins~\cite{hoff1997molecular,spudich2000retinylidene}, cryptochromes~\cite{ahmad1993hy4}, phototropins~\cite{huala1997arabidopsis} and BLUF proteins~\cite{gomelsky2002bluf}. Additionally, we distinguish fluorescent proteins as light-sensitive proteins, which are worth mentioning in the context of light-sensing protein research.

The division of photoactive proteins into families is based on the structure of the chromophore present in their molecules and their mechanism of action. These mechanisms can vary within each protein family. Xanthopsins, phytochromes, and rhodopsins act by photoisomerization of the chromophore. Cryptochromes are activated by electron transfer, BLUF proteins by proton transfer, and phototropins require cysteinyl adduct formation. Unlike the families already named, fluorescent proteins have a different mechanism for responding to light. Inside the protein structure is a fluorophore. In green fluorescent protein (GFP), it consists of three amino acids (serine, tyrosine, glycine) that undergo transformations in the protein environment: cyclization and dehydration, leading to the formation of a fluorescent system of conjugated double bonds~\cite{barondeau2006structural,zhang2006reaction}.

A key feature of photoactive proteins is the presence of a chromophore, which allows them to respond to a specific wavelength of light (Figure~\ref{fig:spectrum}). Upon absorption of a photon by the chromophore molecule, a photochemical reaction takes place, resulting in a signal of conformational change that is propagated to other regions of the protein, inducing specific biochemical functions of the protein, such as participation in signaling pathways. Photoactive proteins often act as molecular switches, returning to a ground state that allows re-excitation (Figure~\ref{fig:photoproteins}).



\begin{figure}[t]
  \centering
  \includegraphics[width=\textwidth]{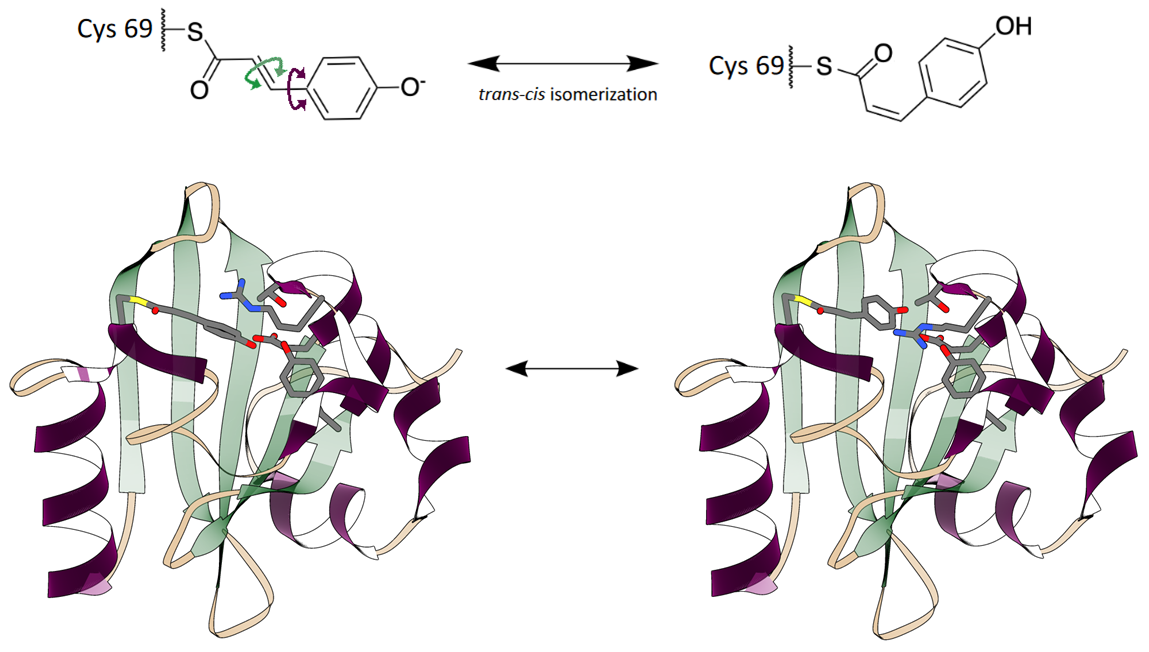}
  \caption{PYP in pG state on the left and pB state on the right (PDB:1TS6)~\cite{ihee2005visualizing}.}
  \label{fig:pyp}
\end{figure}

\subsection{Xanthopsins (PYP and pCA)}
Xanthopsins are a family of light-sensitive proteins. They are characterized by the presence of \textit{para}-coumaric acid (\textit{p}CA) in their structure. This chromophore is attached to the protein by a thiol-ester bond. The photoactive yellow protein (PYP) originally found in the halophilic purple bacterium \textit{Halorhodospira halophila} is a model protein of the xanthopsin family. PYP has a typical $\alpha/\beta$ fold (shown in Figure~\ref{fig:pyp}), which has become the prototype of the PAS domain, which is a key element in biological signaling found in all living organisms, with a central five-stranded $\beta$-sheet and helical segments on either side.

PYP from \textit{H. halophila} is a blue light-sensitive photoreceptor involved in phototaxis with a well-studied reversible photocycle. Light in the range 355--465 nm (Figure~\ref{fig:spectrum}) is absorbed by the ground trans-state chromophore (pG), inducing photoisomerization leading to the formation of an intermediate red-shifted cis-state (pR). Then intramolecular proton transfer occurs, resulting in the formation of a signal cis-blue-shifted state (pB) and transduction of conformational changes in the protein~\cite{van2004photoreceptor,hoff1994measurement,kort1996evidence,xing2022photoactive,hellingwerf2003photoactive}.

The isomerization process in $p$-Coumaric acid occurs by rotation along the double bond. However, studies have reported additional rotation relative to the adjacent single bond, as shown in Figure~\ref{fig:pyp}~\cite{stahl2011involvement,mustalahti2020photoactive}. These rotations cause a change in the architecture of the hydrogen-bonding network around the chromophore and induce a signal for conformational changes throughout the protein. The most important amino acids in this network are Glu46, Tyr42, Thr50, and Arg52, and linking the chromophore to the structure of the protein Cys69 (can be seen in Figure~\ref{fig:pyp} as licorice visualization style colored by atoms)~\cite{sigala2009hydrogen,hellingwerf2003photoactive,xing2022photoactive}. 

\begin{figure}[b]
  \centering
  \includegraphics[width=\textwidth]{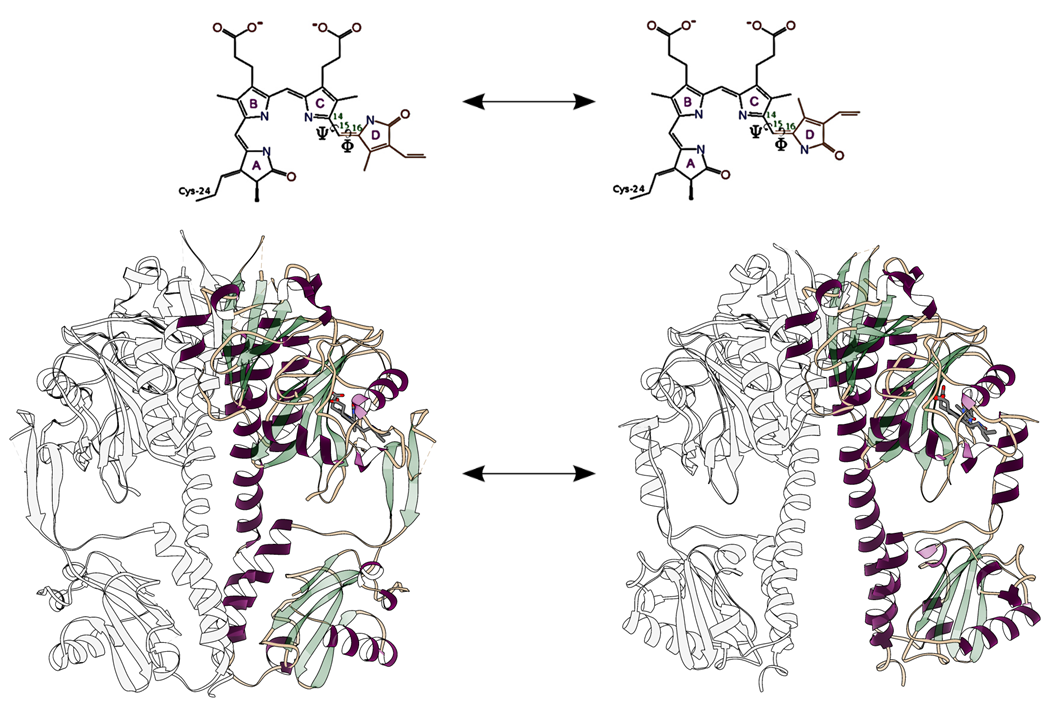}
  \caption{Structural features of the BphP-BV complex. The photoisomerization of biliverdin from the ZZZ state to the ZZE state (top). Bacterial phytochrome in the Pr state on the left and in the Pfr state on the right~\cite{rydzewski2022enhancing}.}
  \label{fig:bphp}
\end{figure}

\subsection{Phytochromes}
Phytochromes are a family of photoreceptors found in the cells of plants, bacteria, and fungi, capable of responding to light or its absence. Their absorption maximum is at wavelengths corresponding to red and far-red light~\cite{legris2019molecular,rydzewski2022enhancing}. Phytochromes such as xanthopsins belong to the Per-Arnt-Sim (PAS) domain protein superfamily, whose name is derived from the three proteins in which it was first identified: Per - period circadian protein, Arnt - aryl hydrocarbon receptor nuclear translocator protein, and Sim - single-minded protein~\cite{moglich2009structure}.

Phytochromes are characterized by the presence of a chromophore molecule belonging to the tetrapyrrole group in their structure. Tetrapyrroles are a class of compounds containing four pyrrole or pyrrole-like rings. The rings are connected linearly or cyclically, but in either case, a double bond pattern is formed that allows the absorption of electromagnetic radiation. They respond to light in the red and far-red ranges of the visible spectrum. In their action as molecular switches, they work between two forms (Figure~\ref{fig:spectrum}): Pfr, which is stimulated by far-red light (705-740 nm), and Pr triggered by red light (650–670 nm)~\cite{rockwell2006phytochrome,li2011phytochrome,halliday2016light,mroginski2011elucidating}.

A representative protein of the phytochromes family is bacterial phytochrome (BphPs), which contains a biliverdin IX$\alpha$ (BV) molecule as a chromofor. The stem module of the BphP photosensor has a three-part region that consists of well-conserved domains: PAS, GAF, and PHY (shown in Figure~\ref{fig:bphp})~\cite{rydzewski2022enhancing,mroginski2011elucidating}.

The light-regulated action of phytochromes is based on the cyclic switching between the Pr and Pfr forms. The transition is triggered by the photoisomerization of ZZZ to ZZE (Pr to Pfr) or ZZE to ZZZ (Pfr to Pr). The photoisomerization of BV, as in the case of PYP and $p$-Coumaric acid, proceeds by rotation relative to the double bond and the single bond located between the C and D pyrrole rings. This induces structural changes within the pocket where the chromophore is located. The signal of conformational changes is then propagated to other regions of the protein. The structural differences between the Pr and Pfr states can be seen in Figure~\ref{fig:bphp}. In response to irradiation, it is possible to observe in the protein a straightening of the helical backbone formed by $\alpha$-helixes derived from the two domains GAF and PHY and a complete change in the architecture of the secondary structure (from $\beta$-sheet to $\alpha$-helix) of a part of the PHY domain called the tongue region~\cite{quail2002phytochrome,takala2014signal,rydzewski2022enhancing,isaksson2021signaling,mroginski2007chromophore}. 

Phytochromes in plant cells are responsible for regulating fundamental physiological processes such as photosynthesis, seed germination, flowering, and phototaxis. In the case of bacterial phytochromes, they control various intracellular processes, including gene expression, protein phosphorylation, and degradation, and their role in the movement of calcium and other ions has also been suggested~\cite{legris2019molecular,oliinyk2017bacterial,schmidt2018structural,kuwasaki2022red,qiao2024sensitive}. 

\subsection{Rhodopsins}
The rhodopsin family encompasses several types of proteins, including visual rhodopsins, which are found in \textit{Eukaryotes} and \textit{Archaea}, as well as ion transport rhodopsins and proton pumps, which are present in \textit{Prokaryotes} and \textit{Archaea}. Additionally, there are sensory rhodopsins involved in signaling pathways, which can be found in all organisms ~\cite{beja2000bacterial,sineshchekov2002two,friedrich2002proteorhodopsin}.

\begin{figure}[b]
  \centering
  \includegraphics[width=0.6\textwidth]{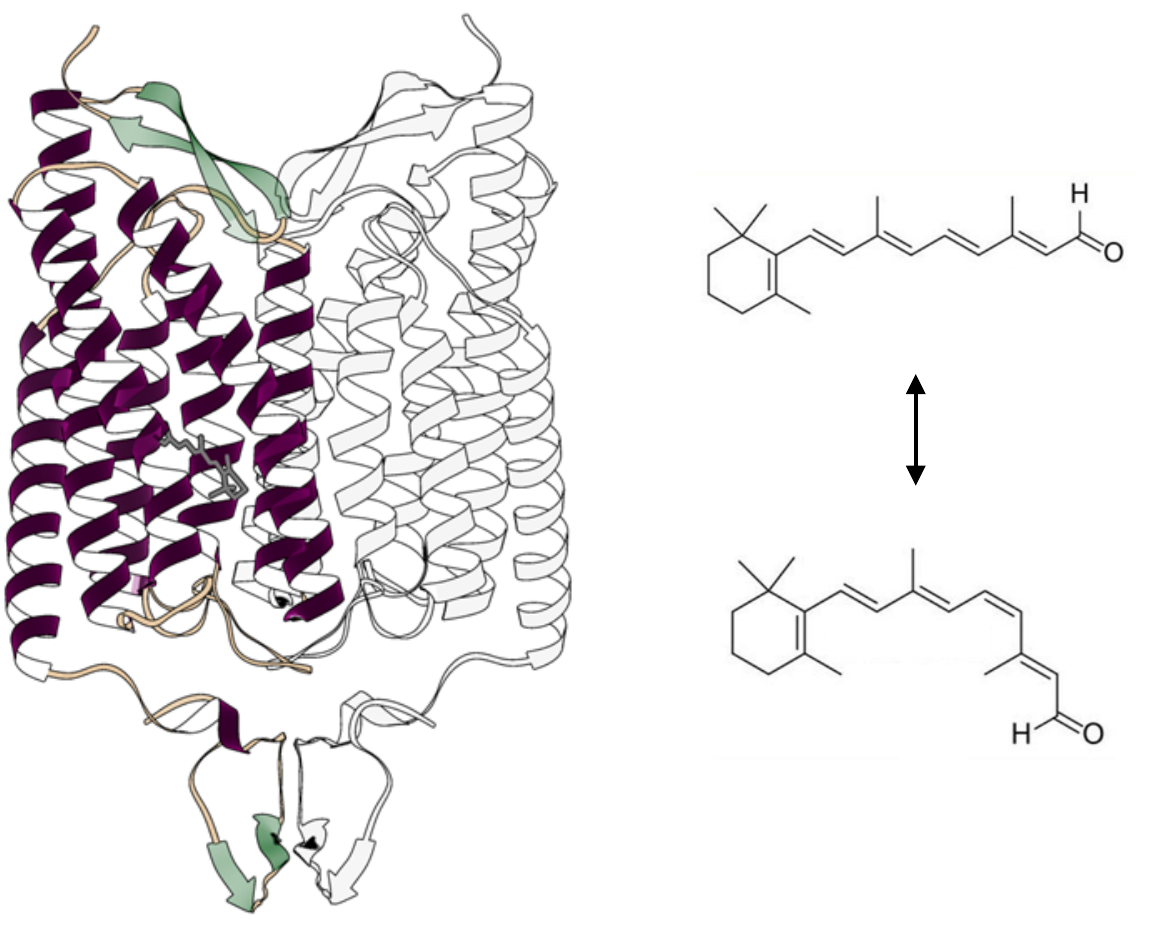}
  \caption{Photoisomerization from all-trans-retinal to 13-cis-retinal at right. At left side, the structure of channelrhodopsin in the closed state with all-trans-retinal (PDB:3UG9)~\cite{kato2012crystal}.}
  \label{fig:chr}
\end{figure}

Eukaryotic rhodopsin is a membrane protein and a member of the G protein-coupled receptors with characteristic seven transmembrane helices that form a region for ligand binding. It consists of the opsin protein and the \textit{11-cis-retinal} chromophore covalently linked to Lys296 (which is located in the seventh transmembrane domain). The chromophore is characterized by the absorption of radiation in the visible light range with a wavelength of approximately 498 nm. After photon absorption, photoisomerization of the chromophore from \textit{11-cis-retinal} to \textit{all-trans-retinal} occurs, leading to conformational changes in the protein and activation of the photoreceptor~\cite{berg2018stryer,caro2015rapid}. 
Eukaryotic rhodopsins, due to their role in vision, are particularly important and well-studied photoactive proteins. Rhodopsins from prokaryotic organisms (and from algae), due to their simpler activation mechanism and lack of coupling to the G protein, have found applications in biotechnology. One type of such rhodopsin is the channelrhodopsin subfamily, which functions as light-gated ion channels. They act as photoreceptors in unicellular green algae and control their phototaxis process~\cite{sineshchekov2002two}. The first proteins of this type to be discovered were Channelrhodopsin-1 (ChR1) and Channelrhodopsin-2 (ChR2) from the model organism \textit{Chlamydomonas reinhardtii}. A recombinant hybrid protein that is a chimera between ChR1 and ChR2 is also being used in biotechnology. All these variants are composed of seven trans-membrane $\alpha$-helices. They contain a chromophore covalently linked to the rest of the protein via a protonated Schiff base, which is \textit{all-trans-retinal}  in the ground state (unlike eukaryotic rhodopsins, where it is \textit{11-cis-retinal}) (Figure~\ref{fig:chr}). 

Channelrhodopsins directly form ion channels and do not require coupling to a G-protein, making cellular depolarization very rapid. Photostimulation is achieved by absorption of blue light with a maximum of 480 nm by the chromophore molecule. This induces a conformational change from \textit{all-trans-retinal} to \textit{13-cis-retinal}. This induces structural changes in the protein, causing the channel to open, allowing ions to pass through. Within several milliseconds, the retinal relaxes to revert to the \textit{all-trans} form, the structural changes propagate, and the channel closes, shutting off the flow of ions~\cite{kato2012crystal,nagel2003channelrhodopsin,bamann2008spectral,govorunova2022kalium}.

Channelrhodopsins, expressed in cells of other organisms, make it possible to control electrical excitability, ion flow, and other cellular processes using light, a method called optogenetics~\cite{zhou2025optogenetics,ovechkina2025advances,charette2025optogenetics,zhou2025advances}.  




\subsection{Fluorescent proteins}
Fluorescent proteins (FPs) have the ability to fluoresce when exposed to light. As most of the known proteins can be fused to FPs, they have become markers for biochemical events that can be observed in living cells. This allows the real-time monitoring of metabolic pathways.

The best known and most widely used example of this type of macromolecule is the green fluorescent protein (GFP) from the jellyfish \textit{Aequorea victoria}~\cite{tsien1998green}. GFP has a characteristic structure of the so-called $\beta$-barrel, formed by several antiparallel $\beta$-sheets arranged in a cylindrical shape (as shown in Figure~\ref{fig:spectrum}). It contains a fluorophore, which is made up of three amino acids (serine, tyrosine, and glycine) that can undergo cyclization and dehydration, leading to the formation of a fluorescing system of conjugated double bonds. Excitation of the fluorophore occurs at 395 nm and 475 nm. In contrast, the emitted wavelength is 509 nm~\cite{prendergast1978chemical,chalfie1995green,yang1996molecular,tsien1998green,remington2011green,chalfie2005green}. 

In GFP, many quantum effects have been explored, including coherent dynamics where electronic and nuclear motions are synchronized after photoexcitation~\cite{cinelli2001coherent}, and photonic entanglement occurring during fluorescence emission~\cite{shi2017generation}.

Due to their potential for various applications, several mutations of GFP (green fluorescent protein) have been developed, including color mutants such as blue fluorescent protein, cyan fluorescent protein, and yellow fluorescent protein. These mutants exhibit a broad ultraviolet absorption band centered around 380 nm, with an emission maximum at approximately 448 nm~\cite{yang1996molecular,brejc1997structural,gaytan2024photoactivatable}. Additionally, unique properties can be observed in fluorescent proteins that contain different chromophores, such as UnaG, which interacts with bilirubin. These properties include a red-shifted emission that occurs above 600 nm or a photoconversion from the green-emitting state to the red-emitting state~\cite{kwon2020bright}. GFPs can be genetically encoded into various organisms~\cite{gest2024molecular} or used as optical switches~\cite{nifosi2024reversibly}.

%



\section{Computational Methods for Studying Photoactive Proteins}
Photoactive proteins are among the most important proteins in the world of living organisms. The understanding of the mechanisms of the activation process that takes place in light-sensitive proteins remains a challenge, as the required temporal and spatial resolution is extremely difficult to achieve experimentally. Therefore, computational studies have been the main source of mechanistic insight into photoreceptor activation. This chapter will provide an overview of computer-based research methods used to study biological systems interacting with light. 

\subsection{Bioinformatics}
Bioinformatics is essential for analyzing protein sequences and extracting valuable information regarding their structure, evolution, and interactions. These tools facilitate sequence alignment, structure prediction, and evolutionary analysis. As a result, they provide critical insights into the properties and mechanisms of action of photoactive proteins~\cite{gauthier2019brief,edwards2003bioinformatics,pavlopoulou2011state}.

Sequence analysis and alignment help identify conserved regions that are often critical for photoactivity, such as chromophore binding sites. They also allow the detection of functional motifs and characteristic domains shared with other photoactive proteins, highlighting evolutionary relationships between protein families. Useful tools for this purpose are BLAST (basic local alignment search tool)~\cite{camacho2009blast+} and Clustal Omega~\cite{sievers2011fast}, which can be used for comparison of protein sequences with known sequences that are available in databases~\cite{horst2009locked}.

Functional annotation is a crucial method in bioinformatics that helps identify post-translational modification sites and interaction regions by detecting functional domains. This can be done using tools such as InterPro~\cite{blum2025interpro} and UniProt~\cite{coudert2023annotation,bateman2024uniprot}. When applied to photoactive proteins, functional annotation can reveal how mutations in the protein sequence might impact its functionality. Additionally, it enables the prediction of specific functions related to photoactivation, such as light absorption and signaling~\cite{kyndt2004photoactive,horst2009locked}.

Evolutionary research tools such as MEGA~\cite{tamura2021mega11} or PhyML~\cite{guindon2010new} are used for phylogenetic analysis and tree construction. They make it possible to track the evolution of photosensitive proteins and identify specific adaptations for interacting with light. They also help to determine the evolutionary relationships between members of photoactive protein families~\cite{khatun2020evolution,xing2022photoactive}.

Predicting the three-dimensional structure of proteins from their sequences is possible with tools such as AlphaFold~\cite{jumper2021highly}, Rosetta~\cite{leaver2011rosetta3}, and the SWISS-MODEL server~\cite{waterhouse2018swiss,bertoni2017modeling}. The main limitation of these programs is their inability to predict light-induced structural changes in proteins. However, they allow one to obtain the ground state structure, and this, together with MD methods, allows us to study light absorption-induced conformational changes, such as chromophore conformational changes and signaling pathways to other parts of the protein~\cite{edwards2003bioinformatics}.

An example of using bioinformatics to analyze photoactive systems has been published in work by Kyndt et al.~\cite{kyndt2004photoactive}, where the authors performed sequence alignment and analysis using BLAST to compare protein sequences (e.g., alignment of bacterial rhodopsins and PYPs to determine conserved regions and functional residues). Additionally, phylogenetic analysis has been conducted to infer evolutionary relationships between phytochrome-related proteins in different bacterial species. Bioinformatics tools also helped to identify and characterize a new family of carotenoid proteins~\cite{bao2017additional,muzzopappa2017paralogs}. Moreover, sequence alignment has enabled the construction of a phylogenetic tree, classification of protein sequences, and prediction of the functional roles of bacteriophytochrome-like proteins in work by Nadella et al. ~\cite{nadella2018identification}. Using bioinformatics, Hasegawa et al.~\cite{hasegawa2020unique} have shown that the genome of cyanobacteria may also contain genes encoding rhodopsins, suggesting that these proteins may have evolved in photoautotrophic lineages.

\subsection{Molecular Dynamics}
\subsubsection{Standard MD Simulations}
Molecular dynamics (MD) simulations have become important for complementing experimental results and achieving insight into complex systems at the microscopic level. In principle, MD is a computer simulation method for analyzing the physical motions of atoms and molecules and the deformation and interaction of molecules over timescales often inaccessible in experiments~\cite{hansson2002molecular,leimkuhler2015molecular,karplus1990molecular}. MD methods are able to simulate the time-dependent behavior of a molecular system. By integrating a set of differential equations derived from classical dynamics, we can obtain the trajectory in the phase space of positions and momenta~\cite{shchukin1981molecular,xiang1994molecular,tuckerman2023statistical}.

In MD simulations, the potential energy function of a system is derived from empirical force fields. The components of the commonly used force fields are parameters such as bond strain (harmonic potential that approximates Hooke's law), angular deflection (harmonic potential for angles between bonds), dihedral torsion (periodic functions for rotation around bonds), and non-bonded interactions (van der Waals: Lennard-Jones potential and electrostatics: Coulomb's law)~\cite{levitt2014birth,karplus2014development,gonzalez2011force,hollingsworth2018molecular}. Examples of popular force fields include CHARMM~\cite{mackerell1998all}, AMBER~\cite{cornell1995second}, OPLS~\cite{jorgensen1996development}, and GROMOS~\cite{oostenbrink2004biomolecular}.

Software such as GROMACS~\cite{abraham2015gromacs}, NAMD~\cite{phillips2005scalable}, and AMBER~\cite{case2020amber2020} are commonly used to study processes such as protein conformational dynamics~\cite{vreede2010predicting,jin2021predicting}, protein-ligand interactions~\cite{baron2013molecular,rydzewski2017ligand}, and intermediate states and transition pathways~\cite{sfriso2013exploration}.

The situation becomes more complicated for photoactive systems, which, in addition to the protein component (modeled via force fields), includes non-standard chromophore molecules that require additional parametrization. This process relies on additional quantum chemical calculations that provide essential information about the molecule, including bond lengths, angles between atoms, dihedral angles, and charge, as well as the connection between the chromophore and the protein component~\cite{breyfogle2023molecular,claridge2018developing,lin2019force,he2022recent}.

\subsubsection{Quantum Calculations}
Quantum calculations play a significant role in the study of photoactive systems by providing detailed insights into the electronic structure and photochemical properties of their chromophores~\cite{gozem2017theory}. One of the most commonly used photoactive compounds is azobenzene. When exposed to light, azobenzene transitions from its trans state to a cis state, followed by thermal relaxation. Andrzej Sokalski et al. have employed a variety of quantum chemical methods to investigate the impact of protonation on this process~\cite{sokalski2001new}. Although this study was not directly related to proteins, the application of quantum methods highlighted that protonation, or the immediate surroundings of a chromophore, significantly influences its dynamics by altering potential energy barriers.

Methods such as time-dependent density functional theory (TD-DFT) and multi-configuration approaches (e.g., CASSCF) allow us to characterize light absorption, excited-state dynamics, and photoinduced reactions at the atomic level. Quantum calculations are essential for understanding the interactions between the chromophore and its protein environment, including charge transfer and polarization effects. They also complement classical techniques that provide a mechanistic understanding of processes such as isomerization, fluorescence, and energy transfer in photoactive proteins~\cite{kubavr2023hybrid,manathunga2022computer}.

TD-DFT simulates the response of the electronic density of a system to external perturbations such as light absorption. It can describe electron transitions between ground and excited states with high accuracy, providing insight into absorption spectra and excited state properties~\cite{tandon2019brief,elstner2003approximate}. TD-DFT is mainly used to analyze transitions, such as singlet and triplet states, which are important for processes such as fluorescence and energy transfer. It also helps to understand how protein environments change the absorption spectrum by predicting the absorption maximum. 

Another approach, complete active space self-consistent field (CASSCF), is a wave function-based method. It takes into account the correlation of electrons in a defined active space. The active space for photoactive systems includes the most important orbitals and electrons that are involved in the photochemical process (e.g., the $\pi$ and $\pi*$ orbitals of the chromophore). In optimizing the wave function, inactive orbitals are assumed to be doubly occupied, while secondary orbitals are considered empty. The CASSCF wave function is obtained by first dividing the space of molecular orbitals (MOs) into three parts: inactive, active, and secondary orbitals. The active orbitals can have any possible occupancy, provided that the overall spin and spatial symmetry of the wave function is maintained~\cite{cardenas2021algorithm,senn2009qm}. Post-Hartee-Fock methods, including both static and dynamic electron 
correlation effects, such as CASPT2 or coupled clusters (CC), have also been employed frequently to model photoactive processes~\cite{gozem2017theory,park2017fly,rao2022photoisomerization,barneschi2023assessment,sirimatayanant2024tuning}.

As quantum calculations are computationally demanding for larger systems, mainly chromophores that exhibit photoactive potential 
have been studied, for instance,  biliverdin~\cite{jozwiak2023molecular,stoll2009structure,polyakov2018modeling}, $p$-Coumaric acid~\cite{semidalas2019structure,garzon2014mechanistic,kumar2015structural}, and retinal~\cite{huix2013assessment,blomgren2005exploring}.

\subsubsection{Hybrid Quantum Mechanics/Molecular Mechanics (QM/MM) Methods}
Quantum mechanics/molecular mechanics (QM/MM) simulations have been essential for modeling photoreceptor proteins, as such large systems cannot be studied effectively using quantum calculations. QM/MM makes it possible to model the chromophore and its surroundings with high precision by applying more expensive quantum methods to describe the chemically active part of the chromophore and its local environment~\cite{andruniow2021qm,levitt2014birth,warshel103theoretical,altun2008quantum}. 


To apply these hybrid methods, it is necessary to divide the considered system into two main regions: the QM region, which includes the atoms directly involved in the molecular process (e.g., the chromophore and the main neighboring amino acids), and the MM region, which is the environment surrounding the QM region (e.g., the other amino acids that make up the protein, the solvent). The QM part is treated with quantum mechanics methods (e.g., Hartree-Fock), DFT, while the classical part is calculated using classical molecular mechanics methods with predefined force fields. To software for QM/MM simulations, we can include ORCA~\cite{neese2012orca}, Gaussian~\cite{gv6}, or ChemShell~\cite{lu2018open}. Programs such as AMBER-Gaussian~\cite{okamoto2011minimal}, GROMACS-CP2K~\cite{kuhne2020cp2k,laino2005efficient}, AMBER/NAMD/GROMACS-TeraChem~\cite{cruzeiro2023terachem,isborn2012electronic} provide interfaces between classical and quantum computing. 

The main challenge in these methods is modeling the interface between QM and MM regions~\cite{groenhof2013introduction,senn2009qm}. As the two regions usually interact directly, it is not possible to write down the total energy of the system as the sum of the component energies~\cite{senn2007qm,lin2005redistributed}. How the two subsystems affect each other and contribute to the total energy of the hybrid system is determined by the coupling between the QM and MM regions. This coupling can involve non-bonded interactions (such as electrostatic and van der Waals) and bonded interactions (such as covalent bonds crossing the QM-MM boundary). The total energy of the QM/MM system as a whole is expressed as~\cite{rossbach2017influence,laino2005efficient}:
\begin{equation}
    E_{total} = E_{QM} + E_{MM} + E_{QM/MM},
\end{equation}
where $E_{QM}$ is the energy of the QM region, calculated using a quantum mechanical method, $E_{MM}$ is the energy of the MM region, calculated using a classical force field, and $E_{QM/MM}$ is the interaction energy between the QM and MM regions~\cite{laino2005efficient,khare2008multiscale}. There are several types of QM/MM coupling schemes that can be distinguished, such as: mechanical, electrostatic and polarizable embedding~\cite{senn2009qm,humeniuk2024multistate}.

In systems simulated using QM/MM, especially when covalent bonds cross the interface, the boundary between the QM and MM regions is critical. Failing to handle this boundary can lead to unphysical results. Usually, non-covalent boundaries are treated by the electrostatic embedding coupling scheme, and van der Waals interactions are handled classically by the MM force field. The treatment of boundaries passing through a covalent bond is more complicated and can be done through several strategies (hydrogen capping, frozen orbitals, localized molecular orbitals, effective core potentials)~\cite{murphy2000frozen,bramley2023application,sun2014exact,watanabe2019quantitative}. 

QM/MM methods can be used to study light-induced processes in biomolecules, which has become a growing area of research in recent years~\cite{tracy2024nonadiabatic,wen2023excited,murphy2023exploring,boulanger2018qm,liang2019nonadiabatic}. PYP has been especially well-studied using QM/MM, including its protonation, hydrogen bonds, and absorption spectrum~\cite{isborn2012electronic,campomanes2021protonation,cruzeiro2021open,tsujimura2022absorption,manathunga2023quantum}. Additionally, QM/MM has been used for channelrhodopsin~\cite{dokukina2019qm}, rhodopsin~\cite{yang2022quantum,di2024fluorescent}, retinal~\cite{punwong2015direct,malakar2024retinal}, and biliverdin~\cite{alavi2018novel,polyakov2018modeling,iijima2018qm,modi2019protonation,santra2024conformational}.

  
  
  
  


\subsubsection{Enhanced Sampling}
MD simulations are essential for studying photoactive systems at a microscopic level; however, their application is often limited by problems with efficient sampling~\cite{valsson2016enhancing,henin2022enhanced,bussi2020using}. This issue is due to rough energy landscapes, with many local minima separated by energy barriers higher than thermal energy, which govern the dynamics of molecules. These limitations can result in an incomplete sampling of conformational states, hindering the analysis and understanding of the functional properties of the system. To accurately characterize the dynamics and function of a system, MD simulations must capture all relevant states of the system. These problems are frequently addressed by using enhanced sampling techniques such as replica exchange~\cite{sugita1999replica}, umbrella sampling~\cite{torrie1977nonphysical}, and metadynamics~\cite{laio2002escaping}. Such enhanced sampling methods are implemented in the PLUMED library, which can be interfaced with many MD codes~\cite{plumed,plumed2019promoting,plumed-tutorials}.

Replica exchange molecular dynamics (REMD) is one of the classical enhanced sampling methods used to explore the conformational space of biomolecular systems efficiently. This is accomplished by running multiple simulations (replicas) of the system under different conditions and periodically replacing configurations between them. This helps avoid local minima and samples a wide region of the energy landscape. For instance, REMD has been used to efficiently sample PYP conformations in work by Vreede et al.~\cite{vreede2008helix} Moreover, REMD simulations of the LOV protein successfully reproduced essential structural changes associated with the light activation process observed experimentally~\cite{ganguly2017glutamine}.

Umbrella sampling (US) is the first enhanced sampling method that employs an external bias potential to improve fluctuations of important degrees of freedom~\cite{torrie1977nonphysical}. In the study by Bondanza et al.~\cite{bondanza2020molecular}, US combined with REMD has been used to study the molecular mechanisms of activation in the orange carotenoid protein. Furthermore, US has been combined by Nottoli et al.~\cite{nottoli2021enhanced} with a polarizable quantum mechanical approach to show the excited state intramolecular proton transfer in solvated 3-hydroxyflavone.

Metadynamics is a computational technique based on adding an external bias potential to the selected degrees of freedom to enhance conformational sampling (similar to US), allowing us to reach the longer simulation timescales and reconstruct the free energy landscapes of complex systems. It has been used for studying photoactive proteins, as light-induced conformational changes and associated energy barriers often pose challenges for conventional MD simulations. For example, metadynamics has been recently used to examine the photoregulation mechanism in the orange carotenoid protein (OCP), bound with a carotenoid chromophore, including conformational transitions of OCP that allow for the carotenoid to translocate and interact with the light-harvesting antenna~\cite{bondanza2020multiple}. Another example is the work by Raucci et al.~\cite{raucci2022enhanced}, where metadynamics has been used to design molecular structures of photoswitches by efficiently exploring their conformational space.


Another enhanced sampling technique used to analyze photoactive proteins is variationally enhanced sampling (VES)~\cite{valsson2014variational}. Recently, VES was used to study the mechanism of phytochrome photoisomerization was investigated, focusing on the biliverdin IX$\alpha$ (BV) - bacteriophytochrome system, particularly the signal transduction pathways and thermal transitions~\cite{rydzewski2022enhancing}.

\section{Conclusion and Future Prospects}
In summary, computational techniques are routinely employed to study photoactive processes. They not only complement experiments but also help to investigate these processes at a microscopic level of detail, which is often unattainable by experimental methods. Using MD simulations---classical, enhanced sampling, or quantum---is but one example of the success of computational techniques applied to studying photoactive proteins. When combined with cryo-EM and spectroscopy, they allow us to decipher the complex molecular pathways of many light-activated processes, such as photosynthesis, phototaxis, and light-dependent gene expression. However, while computational techniques have been successfully used to study photoactive proteins, there is still room for improvement in several areas of method development. 

For example, many photochemical processes occur on femtosecond to picosecond timescales, while associated structural and functional responses (e.g., signal transduction) follow on nanosecond to millisecond timescales. Current methods often struggle to bridge these temporal gaps efficiently. Thus, developing and optimizing methods to detect longer time scales of events occurring in photoactive proteins still constitutes a major challenge~\cite{atkinson1989picosecond,billings2024long}. However, bridging multiple timescales can be achieved by combining enhanced sampling simulations with QM/MM, as recently summarized by Rossetti and Mandelli~\cite{rossetti2024exascale}.



The growing interest in machine learning techniques applied in physics and chemistry could lead to significant advancements in methods used to study photoactive proteins~\cite{li2023machine,zhu2024unsupervised,xu2024ultrafast}. One promising area for improvement in modeling photoactive proteins is the development of neural networks for predicting the relationship between structure and function, as well as generative models for protein and drug design~\cite{unke2021machine,poltavsky2021machine,ovchinnikov2021structure,strokach2022deep,janes2024deep}. Additionally, employing machine learning to uncover reaction coordinates directly from MD simulations to understand the dynamics of complex systems is another promising direction worth exploring in this field~\cite{wang2020machine,sidky2020machine,rydzewski2023manifold,mehdi2024enhanced,gokdemir2025machine}. Another apparent issue comes from the current limitations of standard force fields. Photoactive systems typically consist of a protein and a covalently bound chromophore molecule. This combination presents challenges for modeling non-standard force fields for chromophores, as it necessitates additional parametrization, which is often done using costly quantum calculations. Using machine learning to create non-standard force fields is a significant step toward addressing this issue~\cite{unke2021machine}. We anticipate that these advancements will play a crucial role in enhancing our understanding of photoactive proteins and their associated processes.

The limitations of our approaches are further complicated by our still incomplete understanding of complex photoreactions and multiscale processes. This includes the multistep nature of photocycles, the influence of the protein environment, and the implications of quantum effects. Computational modeling of photodynamic events can be validated against high-quality structural and spectroscopic experimental data. This offers a strong basis for method validation and aids in the interpretation of X-ray data~\cite{weik2022insight}. In summary, it will be beneficial to adopt an interdisciplinary approach consisting of advanced simulation methods, machine learning, and experimental methods.

\begin{acknowledgement}
The research was supported by the National Science Center in Poland (Sonata 2021/43/D/ST4/00920, ``Statistical Learning of Slow Collective Variables from Atomistic Simulations''). J. R. acknowledges funding the Ministry of Science and Higher Education in Poland. W. N. was supported by the IDUB MEMO-BIT project (NCU, Poland).
\end{acknowledgement}

\printbibliography

\end{document}